\documentstyle[preprint,aps]{revtex}
\begin{document}
\draft
%\preprint{cond-mat/9310010}
\title{Magnetic Response of Disordered Ballistic Quantum Dots}
\author{ Yuval Gefen$^{1,2}$, Daniel Braun$^1$, and Gilles Montambaux$^1$}
\address{$^1$ Laboratoire de Physique des Solides, Associ\'e au CNRS,
Universit\'e  Paris--Sud, 91405 Orsay, France\\
$^2$ Department of Condensed Matter Physics, The Weizmann Institute of
Science, Rehovot 76100, Israel}
\date{December 1993}
\maketitle
\begin{abstract}
The weak field average magnetic susceptibility of square shaped mesoscopic
conductors is studied within a semiclassical framework. Long semiclassical
trajectories are strongly affected by static disorder and differ sharply
from those of clean systems. They give rise to a large linear paramagnetic
susceptibility which is disorder independent and in agreement with recent
experiments. The crossover field to a nonlinear susceptibility is discussed.\\
\end{abstract}
%\newpage

%\section{Introduction}
Recent developements in the theory of thermodynamics on the mesoscopic scale
have brought to the focus of attention the question of the average magnetic
susceptibility of an isolated mesoscopic counductor, $\langle\chi
(H)\rangle$. It has been shown that as the number of electrons in such a
system is independent of the value of the applied magnetic field, one has to
evaluate this quantity within the canonical ensemble. Upon energy averaging
(or averaging over system size), one may discard oscillatory grandcanonical
contributions, obtaining
\begin{equation}
\langle\chi (H)\rangle=-\frac{\Delta}{2}\frac{\partial^2}{\partial H^2}
\langle\delta N^2(H)\rangle\big|_{\overline{\mu}}\,,
\label{ki1}
\end{equation}
where $\langle\delta N^2\rangle$ is the typical (field dependent) sample to
sample fluctuation in the number of levels below some effective chemical
potential $\overline{\mu}$ \cite{AGI}. Efforts have been concentrated on the
calculation  of $\langle\chi\rangle$ in the diffusive regime, but have been
extended as well to the limit of clean systems (where the only scattering is
due to the sample's boundaries)
\cite{Thomas,Dutch,Oppen,Chaos,Orsay,Ovchinikov}. With the regime
\begin{equation}
l_{el}>L \label{bal}\end{equation}
made accessible experimentally \cite{Levy}, the belief is that such
theoretical studies may be employed to interpret the experimental data. Here
$l_{el}$ is the elastic mean free path, while $L$ is the system size.\\

{}From another point of view, it has become clear that the association of all
systems satisfying Eq.~(\ref{bal}) (perfectly clean or containing disorder)
with a ubiquitous ballistic regime is too naive \cite{Sivan,Altland}. Let us
consider a finite disordered system of an integrable geometry (e.~g.~a
square). For disorder weaker than that which gives rise to diffusive
behaviour, one has to compare not only $L$ with $l_{el}$, but also the
average level spacing, $\Delta$, with the inverse elastic mean free time
$\hbar/\tau_{el}$. We may now distinguish between two weak disorder regimes
satisfying Eq.~(\ref{bal}): the {\em ballistic regime} (
$\hbar/\tau_{el}>\Delta$) and the {\em clean} (or {\em perturbative}) {\em
regime} ( $\hbar/\tau_{el}<\Delta$ or, equivalently, $l_{el}>L(k_FL)^{d-1}$.
This regime is, presently, experimentally inaccessible.). For the latter, as
disorder constitutes only a small pertubation on top of the clean spectrum,
one may argue that thermodynamic quantities are practically those of a
perfectly clean system (of the same geometry). This is, however, not the
case for the former, ballistic regime. It has been shown recently that
elastic mixing of the levels give rise to non--trivial level correlations
\cite{Altland} (at $H=0$). It is therefore natural to ask how thermodynamic
quantities, notably the magnetic susceptibility, behave in this regime. \\

In the present work we address the problem of the average weak field
susceptibility of (integrable) disordered ballistic quantum dots. Employing
a semiclassical picture \cite{AIS} we find contributions to
$\langle\chi (H)\rangle$ which arise  from long trajectories that are
sensitive to elastic scattering. However, due to subtle cancellations that
occur within this framework, contributions to the zero field susceptibility
do not depend on disorder \cite{fn1}. We find that this susceptibility is
given by
\begin{equation}
\chi(H=0)=+|\chi_L|\alpha k_FL\label{ki2}\,.
\end{equation}
It is paramagnetic and includes an enhancement factor $\alpha k_FL$ with
respect to  the Landau susceptibility, where the numerical factor $\alpha$
is estimated below. Our results are in qualitative agreement with the
experimental data of Ref.~\cite{Levy}. Our approach is closely related to
the de Gennes Tinkham ``method of trajectories'', and some of our results
bear close formal similarity to a method, discussed by Beenakker and van
Houten, of treating weak localization corrections in restricted geometries
\cite{Gennes,fn1}.\\

Consider a possible multiply reflected semiclassical trajectory within our
square geometry. The same trajectory may be described within an
extended zone scheme (Fig.~1a), where boundary scattering does not occur.
We next consider a uniform perpendicular magnetic field in the
$z$--direction. Within the Landau  gauge the vector potential is ${\bf A}=-H
y \hat{\bf x}$. The vector potential in the extended zone scheme is shown
schematically in Fig.~1b. According to this scheme, the electron is moving
in a staggered field  \cite{fn2}, which may be written as sum of its Fourier
components
\begin{eqnarray}
{\bf A}=-\frac{16 H L \hat{\bf
x}}{\pi^3}\sum_{m,n=0}^\infty&&\frac{(-1)^n}{(2 m+1)(2 n+1)^2}\sin((2
m+1)\frac{\pi x}{L})\nonumber\\
&&\sin((2 n+1)\frac{\pi y}{L})\,. \label{A}
\end{eqnarray}\\

Imagine now a random walk trajectory (in the extended zone scheme) of total
length $\cal L$, starting from point $(x_0,y_0)$, and consisting of ${\cal
L}/l_{el}$ uncorrelated segments of length $l_{el}$ each. The end point of
this trajectory may be mapped back onto the reduced zone scheme. For long
trajectories (${\cal L}>l_{el}$) we assume that ergodicity holds and that
the image of the end point in the reduced zone scheme is uniformly
distributed. In the absence of an applied field, the probability of
returning to the origin at time $t$ is given by (see e.~g.~ \cite{AIS})
\begin{equation}
p(t)\equiv p_0=2\frac{1}{(2\pi\hbar)^2}\label{p0}\,,
\end{equation}
where the factor $2$ in front represents an enhancement due to an
interference of a trajectory with its time reversed image (a ``Cooperon
contribution''). The two--level spectral correlation function, $Y_2(\Delta
E)$, is given then by the fourier transform of $|t|p(t)$ \cite{AIS}, and
finally the level fluctuation is given by
\begin{eqnarray}
\langle\delta N^2(\Delta E)\rangle&=&2\int_0^{\Delta E}(\Delta
E-s)Y_2(s)\,ds\nonumber\\
&=&4\int_0^\infty\,dt\frac{\hbar^2 p(t)}{t}(1-\cos(\frac{\Delta E t}{\hbar}))
\label{dN2}\,.
\end{eqnarray}
Our goal is to find the field dependence of $p(t)$, hence $\langle\delta
N^2\rangle$, then employ Eq.~(\ref{ki1}). We first note that considering
long trajectories, i.~e.~ time scales larger than the elastic mean free time
$\tau_{el}$, allows us to evaluate reliably energy windows such that $\Delta
E<\frac{\hbar}{\tau_{el}}$. This, in turn, may be employed to evaluate the weak
field susceptibility. (In other words, long trajectories affect the small
energy contribution of the level--level correlation function to $\langle
\delta N^2\rangle$ \cite{fn1,Imry91}).
Estimates of the field scales for which the present treatment is valid are
given below. Upon application of a magnetic field there is an extra phase
that may be associated with the amplitude of the $j$th returning
trajectory, $\phi_j=\frac{2 \pi}{\phi_0}\int_{(j)} {\bf A}\cdot\,d{\bf x}$,
where the integral is taken along the $j$th trajectory. To calculate $p(t)$
one has to take the square modulus of the sum over all trajectories
corresponding to time $t$. Upon disorder averaging, only terms that may be
represented as a product of an amplitude with its complex conjugate or with
the complex conjugate of its reversed survive \cite{AIS,Bergmann84}. The
factor 2 (Eq.(\ref{p0})) is now to be replaced by a factor $(1+\cos
2\phi_j)$ to be averaged over all trajectories ${j}$ \cite{fn7} . We note
that following the disorder averaging, the dependence on random (trajectory
specific) phases disappears. Our remaining task is to evaluate the
distribution function of $2\phi_j$. We first assume that the magnetic phases
(contributions from line integrals) accumulated at different steps of size
$l_{el}$ of a trajectory are independent and identically distributed, with a
distribution function $P_{l_{el}}(\tilde{\phi})$. The average over $\cos
2\phi_j$ is then given by \cite{fn2}
\begin{eqnarray}
&&{\rm Re}\left[\int_{-\infty}^\infty
P_{l_{el}}(\tilde{\phi})e^{i2\tilde{\phi}}\,d(\tilde{\phi})\right]^{{\cal
L}/l_{el}}\nonumber\\
&=&{\rm Re}\left[\frac{1}{2\pi}\int_{-\infty}^\infty\int_{-\pi}^{\pi}
(\delta(\tilde{\phi}- \phi(\theta))e^{i2\tilde{\phi}}\,d\theta)\,d(
\tilde{\phi})\right]^{{\cal L}/l_{el}}\nonumber\\
&=&{\rm
Re}\left[\int_{-\pi}^{\pi}\frac{d\theta}{2\pi}e^{i2\phi(\theta)}\right]^{{\cal
L}/l_{el}}\equiv{\rm Re}\,(\zeta(H))^{{\cal L}/l_{el}}\label{Re}\,,
\end{eqnarray}
where $\theta$, the scattering angle at a given step (cf.~Fig.~1a), is assumed
to be uniformly distributed (a consequence of isotropic scattering). The
function $2\phi(\theta)$ (associated with a segment of length $l_{el}$) is
evaluated by inserting Eq.~(\ref{A}) into the expression for the line integral.
Approximating $A_x(x,y)$ by the $m=n=0$ terms in Eq.~(\ref{A}), we obtain
\begin{eqnarray}
2\phi(\theta)&=&-2\cdot 2\pi\frac{8 H L}{\pi^3\phi_0}
\int_0^{l_{el}}ds\cos\theta\nonumber\\
&&\big[\cos\left(\frac{\pi}{L}(x_0-y_0)+\frac{\pi s}{L}(\cos\theta-\sin\theta)
\right)\nonumber\\
&&-\cos\left(\frac{\pi}{L}(x_0+y_0)+\frac{\pi s}{L}(\cos\theta+\sin\theta)
\right)\big]\label{AA}\,,
\end{eqnarray}
where $(x_0,y_0)$, $(x_0+l_{el}\cos\theta,y_0+l_{el}\sin\theta)$ are the
endpoints of the segment at hand. We next write
\begin{equation} p(t)=\frac{1}{2}p_0 (1+{\rm Re}\,\zeta(H)^{{\cal L}/l_{el}})
\nonumber
\end{equation}
Accounting only for the field sensitive (Cooperon like) contribution $\Delta
p(t)$, writing ${\cal L}=v_Ft$, $l_{el}=v_F\tau$ and introducing a cutoff
factor \cite{fn3}, $\gamma$, we obtain
\begin{equation}
\Delta p(t)=\frac{1}{2}p_0\zeta^{\frac{t}{\tau}}e^{-\frac{\gamma t}{\hbar}}
\label{deltap}\,.
\end{equation}
Substituting this into Eq.~(\ref{dN2}), we obtain that the flux sensitive
part of $\delta N^2(\Delta E)$ is given by \cite{fn4}
\begin{equation}
\delta N^2(\Delta E)=p_0\hbar^2\ln\frac{(\Delta
E)^2+(\frac{\hbar}{\tau_{el}}\ln\frac{1}{\zeta}+\gamma)^2}
{(\frac{\hbar}{\tau_{el}}\ln\frac{1}{\zeta}+\gamma)^2}\label{dN2.2}
\end{equation}\\

To evaluate the very weak field behaviour we may expand
\begin{equation} \zeta(H)\simeq
1-\int_{-\pi}^{\pi}\frac{d\theta}{4\pi}(2\phi(\theta))^2\equiv
1-c\frac{H^2L^4}{\phi_0^2}\label{ki(H)}\,.
\end{equation}
Averaging over $x_0,y_0$ we obtain
\begin{equation}
c=\frac{32}{\pi^5}\int_{-l_{el}/L}^{l_{el}/L}ds\int_{-l_{el}/L}^{l_{el}/L}ds'
\int_{-\pi}^{\pi}d\theta\,\cos(\pi\sqrt{2}(s'-s)\sin\theta)\label{c1}\,,
\end{equation}
which, following a few straightforward steps, yields
\begin{equation}
c=\frac{512}{\pi^4}\int_0^{l_{el}/L}du\, J_0(\sqrt{2}\pi u)(\frac{l_{el}}{L}
-u)\label{c2}\,,
\end{equation} where $J_0$ is a Bessel function.
As $\frac{l_{el}}{L}\gg 1$, this integral may be approximated by
\begin{equation}
c\simeq\frac{256\sqrt{2}}{\pi^5}\frac{l_{el}}{L}\label{c3}
\end{equation}
The most remarkable feature of Eq.~(\ref{c3}) is the proportionality of
$c$ to $l_{el}$. The main contribution to the quadratic correction $cH^2$
comes from segments oriented at angles $\theta\simeq\pm\pi/4$, for which
$\langle\phi^2\rangle\propto l_{el}^2$. However, the angular width around
$\theta=\pm\pi/4$ corresponding to such  exceedingly large contributions is
$\sim\frac{L}{l_{el}}$, rendering $c\sim l_{el}$. It follows then that
$\zeta^{\frac{{\cal L}}{l_{el}}}\simeq 1-\frac{{\cal
L}}{l_{el}}c\frac{H^2L^4}{\phi_0^2}$,
rendering the
quadratic term, hence  the zero field susceptibility, {\em disorder
independent}.
Employing Eqs.~(\ref{ki1}) and (\ref{dN2.2}), and considering an energy window
$\Delta E\simeq \frac{\hbar}{\tau_{el}}\gg\Delta$, we obtain
\begin{eqnarray}
\langle\chi(H=0)\rangle&=&\frac{\Delta}{\gamma}p_0\hbar^3\frac{2c}{\tau_{el}}
\frac{L^4}{\phi_0^2}=\frac{1536\sqrt{2}}{\pi^{8}}\frac{\Delta}{\gamma}
|\chi_L|k_FL\nonumber\\
\simeq
&&0.23\frac{\Delta}{\gamma}|\chi_L|k_FL\,,\label{ki0}
\end{eqnarray}
where for spinless electrons $\chi_L=-\frac{e^2L^2}{24\pi m c^2}$. For
$\frac{1}{2}$ spin electrons $\chi$, as well as $|\chi_L|$, should be
multiplied by 2 \cite{Dupuis}.
 The range for which this very weak field analysis may be employed (i.e.~the
regime of linear susceptibility,  eq.~(\ref{ki0})) is found by ({\sl i})
examining the
range of validity of Eq.~(\ref{ki(H)}), i.e., by requiring that $\zeta$ may
be expanded to quadratic order in $H^2$ and ({\sl ii}) by requiring that the
field dependent correction in Eq.~(\ref{dN2.2}) be smaller than $\gamma$!
These conditions read
\begin{equation}\begin{array}{lcr}
H<\frac{\phi_0}{Ll_{el}}&(\mbox{flux in the sample}
<\frac{L}{l_{el}}\phi_0) \cite{fn5}& ({\sl i})\\
H<\phi_0\sqrt{\frac{\gamma\tau_{el}}{l_{el}L^3}}&&({\sl
ii})\end{array}\label{H1}
\end{equation}
Typically, inequality (\ref{H1}{\sl ii}) is stricter, and yields the
dependence of the crossover field on both the elastic and inelastic scattering.
 Beyond this range  the
susceptibility is not constant, and may even change sign (cf.~Fig.~2). We next
recall that
our semiclassical considerations apply to energy intervals (close to
$\epsilon_F$) smaller than $\frac{\hbar}{\tau_{el}}$.  We now find the largest
field for which this picture may be employed to evaluate the total
$H$--dependence of $\langle \delta N^2\rangle$ (hence $\chi$).  The condition
is that $\langle(2\phi(\theta))^2\rangle<1$, yielding
\begin{equation}
H<\frac{\phi_0}{\sqrt{l_{el} L^3}}\label{H2}\,.
\end{equation}
\\

To obtain contributions from long (${\cal  L}>l_{el}$) trajectories, we
require that the dephasing length ($l_\phi$) satisfies $l_\phi>l_{el}$. In
this case $\gamma\simeq\frac{\hbar D}{l_\phi^2}$, hence $\chi/|\chi_L|$
scales as $l_\phi^2/Ll_{el}$.
The magnetic susceptibility of two--dimensional ballistic dots (squares) has
recently been measured  by Levy et al.~\cite{Levy}. They have obtained
$\frac{\chi(H=0)}{|\chi_L|}\simeq k_FL\simeq 100$. In their experiments
$l_{el} \simeq L-2 L$, $l_\phi\simeq 3L-10L$ .
Eq.(\ref{ki0}), pushed to the limit of its validity, yields ($l_{el}=1.5 L$,
$l_\phi=8L$) $\chi\simeq 250
|\chi_L|$, in rough agreement with the experiment. Evidently, a more
detailed comparison with our theory (e.~g., measurements of the
crossover field from linear susceptibility, which is sensitive to disorder)
is needed before one may confirm
the validity of our picture. \\

To summarize, the central point of our analysis was based on the observation
that when the condition $l_\phi\gg l_{el}$ is met,
  the weak field behaviour carries the signature of the {\em
long}, disorder dependent trajectories. It is {\em possible} that the
contribution calculated in Refs. \cite{Chaos,Orsay} should be added on top
of the present result \cite{fn8}.
It is the dependence on $l_{el}$ that
cancels out which gives rise to the disorder independent  result,
Eq.~(\ref{ki0}), in sharp contrast with, e.g., the results expected in the
Aharonov--Bohm case. The scale of the field for which the susceptiblity
decreases  may be disorder dependent (Eq.~(\ref{H1})) \cite{fn6}.

{\it Acknowledgments: }
We have benefitted from useful discussions with A.~Alt\-land, N.~Argaman,
R.~A.~Jalabert, K.~Richter, and D.~Ullmo. In particular we are grateful to
D.~Mukamel for pointing out to us the usefulness of the expansion
Eq.~(\ref{A})  in the present context.
Y.~G.~acknowledges  the
hospitality of H.~Bouchiat and G.~Montambaux in Orsay. This research
was supported in part by the German--Israel Foundation (GIF), the
U.~S.~--Israel Binational Science Foundation (BSF), the Claussen
Stiftung, and EC Science program no. SCC--CT90--0020.\\

\begin{figure}
FIG.~1.a.
A multiply reflected semiclassical trajectory (thick line), and its image
within the extended zone scheme. A dot denotes the location of a scatterer.

\end{figure}
\begin{figure}
FIG.~1.b. Vector potential within the extended  zone scheme (arrow lengths
are proportional to $|A_x|$). Alternating directions of the staggered field
are indicated.
\end{figure}
\begin{figure}
FIG.~2. The field dependent susceptibility calculated numerically
(Eqs.~(\ref{ki1}) and (\ref{dN2.2})) with $x_0=y_0=0$ for
$H<\frac{\phi_0}{\sqrt{l_{el} L^3}}$. Here $l_{el}/L=50.3$. Note that
$\langle\chi(H)\rangle$ changes sign for $H>\frac{\phi_0}{Ll_{el}}$. These
oscillations will be partially smeared out due to fluctuations in the value
of $l_{el}$.
\end{figure}

\begin{references}
\bibitem{fn2} We treat these trajectories as consisting of an integer number
of equal length steps. This ``discretization'' of the paths introduces small
oscillatroy terms on top of our results (e.g.~Eq.(\ref{c2})), which will be
smeared out upon assigning a continuous distribution to $l_{el}$.
\bibitem{fn7} As we consider bilinear products of the amplitudes the
dependence on Maslov indices disappears.
\bibitem{fn3} The factor $\gamma$ may represent inelastic scattering.
However, even when the inelastic time is large, $\gamma$ should be of the
order of $\Delta$, accounting for the fact that for $t>\hbar/\Delta$ our
semiclassical picture ceases to be valid \cite{Berry,AS}.
\bibitem{fn4} Strictly speaking Eq.~(\ref{deltap}) holds only for
$t\gg\tau_{el}$. The error in substituting Eq.~(\ref{deltap}) into
Eq.~(\ref{dN2}) can be shown to be small.
\bibitem{fn5} The analogue of this condition in the diffusive regime is
$\phi<\frac{1}{\sqrt{g}}\phi_0$, where $g$ is the dimensionless conductance
(cf.~\cite{AGIM}).
\bibitem{fn8} When dealing with two--dimensional electron gas systems, one
may encounter situations where the transport mean free path is larger than
$L$ (i.~e., the system is ballistic), while $l_{el}\ll L$. This may imply that
invoking a semiclassical description applied to a clean system (employing
continuous families of periodic trajectories, cf.~Refs. \cite{Chaos,Orsay})
may not be adequate.
\bibitem{fn6} We propose the following conjecture concerning  the
generalization  of this
result to clean systems of non--integrable geometry. In such geometries
ergodicity is achieved  for trajectories ${\cal L}>L$. One may be tempted
then to employ the results of the preceeding analysis, with the
identification $L\sim l_{el}$. That would imply that on the axis
$\frac{L}{l_{el}}$, generic quantum chaotic systems  are basically represented
by one  point.
Evidently one may produce non--integrable systems for which the effective
$\frac{L}{l_{el}}\ll 1$  (e.g.~a sequence with one of its corners a bit
rounded).
\end{references}
\end{document}